\documentclass[12pt]{article}
\usepackage{graphicx}
\begin{document}

\newpage
\hrule
\vspace{5cm} 
\centerline{\Large \bf The p=0 condensate is a myth}  

\vspace{2cm}
\bigskip
\centerline{\bf Yatendra S. Jain}
\smallskip
\centerline{Department of Physics,} 

\centerline{North-Eastern Hill University, Shillong-793 022, India}   
\vspace{2.5cm}
\bigskip
\begin{abstract} 
Analyzing some of the basic aspects of the dynamics of two 
bosons (interacting through a central force) and their importance 
in determining the ground state of a system like liquid $^4He$, 
it is unequivocally concluded that our conventional belief in 
the existence of $p=0$ condensate in the superfluid state of 
such systems [including the state of Bose Einstein condensate
(BEC) of trapped dilute gases] is a myth. 

\end{abstract}
\vspace{1.25cm}

\noindent
e-address: ysjain@email.com

\vspace{1.25cm}

\noindent 
\copyright by author
\smallskip
\hrule

\newpage

\bigskip 
\centerline {\bf 1. Introduction}

\bigskip
Over the last seven decades, numerous efforts have been made 
to develop the microscopic theory of a {\it system of interacting 
bosons} (SIB) like liquid $^4He$ by using different mathematical 
tools such as variational principle or perturbation technique 
clubbed with the {\it most popular presumption} that there exists 
zero momentum ($p=0$) condensate in the single particle state in its 
superfluid phase 
(see recent articles [1-5] for other important reviews and research papers 
published on the subject after Bogoliubov's theory [6] of weakly 
interacting bosons).   Nevertheless it is clear that our microscopic 
understanding of superfluidity of a SIB based on the said $p=0$ 
condensate is still open to questions.  This is corroborated by the fact 
that more recently [7],  efforts have also been made to develop an 
alternative theory by introducing coherent pair state ({\it analogous to the 
Cooper condensate in a Fermi liquid with attraction between fermions}) along 
with the existence of $p=0$ condensate.  Guided by these facts we tried 
to develop another alternative theory [4,5] of such a SIB by obtaining 
solutions of $N$-particle Schr\"{o}dinger equation {\it without making any 
ad hoc assumption such as the existence of $p=0$ condensate} in its 
superfluid state.  Our theory concludes that: (i) each particle in the system 
represents a pair of particles moving with equal and opposite momenta 
({\bf q}, -{\bf q}) with respect to their {\it center of mass} 
(CM) which moves in the laboratory frame with momentum {\bf K} 
[see Section 2.1 below for detailed meaning of {\bf q} and 
{\bf K}], (ii) the onset of the superfluid transition 
represents the {\it Bose Einstein condensation} (BEC) 
of these particles (as the representatives 
of the pair) in a state of $K=0$ and $q=q_o = \pi/d$ and the 
system moves from its disordered state in the phase space defined 
by $\Delta\phi \ge 2\pi$ (with $\Delta\phi$ representing the 
relative phase position of two neighboring particles) to an ordered state 
defined by $\Delta\phi = 2n\pi$ ($n =$ 1, 2, 3, ...) shifting  
particles from their random locations (in position space) 
in the high temperature phase to a close packed arrangement 
of their {\it wave packets} in low temperature phase keeping every 
two neighboring particles at an identically equal distance 
$d = ({\rm V}/N)^{1/3}$ (with V being the total volume occupied 
by $N$ particles); it allows particles to move only in the 
order of their locations ({\it obviously with no collision or 
relative motion leading to zero viscosity}). The theory explains 
superfluidity 
and related properties of liquid $^4He$ to a very good accuracy 
at quantitative level [4, 8].  Based on {\it pair of particles 
basis} (PPB), our theory differs significantly 
from conventional theories based on {\it single 
particle basis} (SPB).  For example, the {\it ground state} 
(G-state) of the system, whose correct understanding 
has a key role in revealing the physical behavior of the 
system, is found to have following important differences, 
-identified from the results of SPB and PPB. 

\bigskip
\noindent 
{\bf G-state(SPB)} : Different number of atoms have different 
momenta with certain percentage ({\it e.g.}, $\approx 10\%$ $^4He$ 
atoms in superfluid $^4He$) occupying the state of $p=0$.

\bigskip
\noindent 
{\bf G-state(PPB)} : Each of the $N(N-1)/2$ pairs, that we can count by 
way of choosing any two of $N$ particles, has CM 
momentum $|{\bf K}|=0$  and each atom in the pair, having localized 
position within the cavity of size $d$ left by its neighbors for its 
exclusive occupancy, has $q_o =\pi/d$ as the least possible value of 
$q$ (with $2{\bf q} = {\bf  k}$ being the relative momentum of two 
particles).

\bigskip
The momentum distribution of particles in G-state(SPB) and G-state(PPB) [5]
of a SIB and that of a {\it system of non-interacting bosons} (SNIB) is 
depicted in Fig.1 for a better understanding of the difference between 
these G-states.  It is natural that the results of only one (either SPB 
or PPB) can be correct and to this effect the following analysis 
unequivocally reveals that only G-state(PPB) concluded by us [4,5] 
represents the true G-state of a SIB.    

\bigskip
\centerline{\bf 2. Dynamics of two particles and G-state of a SIB}

\bigskip
\noindent
{\bf 2.1. Values of K and q in the G-state}

\bigskip
\noindent
{\bf (a).} It is well known that the dynamics of two particles 
({\it say}, P1 and P2 interacting through a two body central 
force), moving with momenta ${\bf k}_1$ and ${\bf k}_2$ in 
the laboratory frame, can always be described in terms of their 
relative momentum ${\bf k} = 2{\bf q} = {\bf k}_2 - {\bf k}_1$, 
and CM momentum ${\bf K } = {\bf k}_1 + {\bf k}_2$, as we have 
$${\bf k}_1 = -{\bf q} + {\bf K}/2, \eqno(1)$$

\noindent
and 
$${\bf k}_2 = {\bf q} + {\bf K}/2. \eqno(2)$$ 
\noindent
{\bf (b).}  Since CM motion of P1 and P2 {\it does not encounter inter-particle 
interaction}, it represents a freely moving body of mass $2m$ implying 
that the G-state of the pair would invariably have $|{\bf K}|=0$.  

\bigskip
\noindent
{\bf (c).}  With $|{\bf K}| = 0$ in Eqns.(1 and 2), we are left with 
$|{\bf k}_1| = q$ and $|{\bf k}_2| = q$  as the residual momenta of 
P1 and P2 whose least possible value ({\it say}, $q_o$) in their 
G-state would be nothing but the momentum of their zero-point 
motion.  

\bigskip
\noindent
Since 2.1(a-c) can be applied to any two 
particles that we pick up in the system, all the $N(N-1)/2$ pairs 
in the G-state of the system have to have $|{\bf K = 0}|$ which 
implies that all the $N(N-1)/2$ CM points (a CM point is the mid point of the 
line joining the centers of mass of two chosen particles) of the  
pairs get fixed in position space which further implies that the 
position of each particle also gets fixed [9], -of course, with certain 
amount of position and momentum uncertainty due to wave particle 
duality.

\bigskip
\noindent
{\bf 2.2.  G-state(SPB) does not ensure $K=0$ for every two particles}

\bigskip
\noindent
{\bf (a).} We note that the energy of the pair is given by 
$$E(2) = \frac{\hbar^2}{2m}(k_1^2+k_2^2) = 
\frac{\hbar^2}{4m}(k^2+K^2) = \frac{\hbar^2}{4m}(4q^2+K^2) 
\eqno(3) $$
\noindent
which implies that it has its minimum value only if $|{\bf q}|$ and 
$|{\bf K}|$ have their minimum values ($|{\bf q}| = q_o$ and 
$|{\bf K}| = 0$).  However, two particles in the G-state(SPB) do 
not have $|{\bf K}| = 0$ except for the particles having {\it equal 
and opposite} ${\bf q}$ ({\it i.e.},  ${\bf k}_1 = -{\bf k}_2 = 
{\bf q}$) and even for this case SPB theories (as shown in Section 2.3, 
below) do not prescribe any condition to ensure minimum value for 
$|{\bf q}|$.  To this effect, it may be noted that during the process 
through which a SIB reaches its G-state, {\it changes in} {\bf q} 
{\it can be totally independent of the changes in} {\bf K}.   

\bigskip
As such we find that the energy of G-state(SPB) [{\it say}, $E_o$(SPB)] 
is not ensured for its minimum value and this is evident from the fact 
that even the single main term of contribution to per particle $E_o$(SPB) 
[1, 10, 11], {\it i.e.}, $4\pi a\hbar^2/mv = ah^2/\pi md^3$ 
(with $v = d^3$ being the average volume per particle and $d$ the 
inter-particle distance), is higher than that $(h^2/8md^2)$ in 
G-state(PPB) [4,5] by a factor of $(8a/\pi d)$ which for liquid 
helium falls around 2.  

\bigskip
\noindent
{\bf (b).} Interestingly, when $|{\bf K}|$ = 0 is 
applied to all pairs of particles in G-state(SPB) to 
minimize $E_o$(SPB) for the energy of their CM motions, 
we immediately find that particles in each pair have 
{\it equal and opposite} $q$ whose minimum value can not 
be less than $q_o = \pi/d$ (see Section 2.3(d) below).  

\bigskip
\noindent
{\bf 2.3.  G-state (SPB) does not ensure minimum $|q|$} 

\bigskip
\noindent
{\bf (a).} Following the statement of shape independent 
approximation (Huang [11], p.279, Section 13.2 of this book), 
``at low energies the potential acts as if it were a {\it hard sphere} 
potential of diameter $a$'' where $a$ represents the s-wave 
scattering length, it is clear that the volume occupied 
exclusively by a particle can not be smaller than $a^3$.

\bigskip
\noindent
{\bf (b).} As found experimentally 
by Grisenti et al. [12] (who also report theoretical $a$ for 
a comparison), $^4He$ atoms have $a \approx 100\AA$ 
at 1 mK energy, while they have $a \approx 3\AA$ at $\approx 1$K 
energy [10]. 
This fact indicates that $a$ of the particles of a 
milli-Kelvin energy is about 30 times larger than that 
of a Kelvin energy.

\bigskip
\noindent
{\bf (c).} In what follows, atoms of different energy/
momentum in G-state(SPB) exclusively occupy different volumes that 
differ significantly.  For example $a^3$ volume occupied by a $^4He$ 
atom of 1 mK energy is estimated to be about 27,000 times that 
of 1 K energy and it could be even more for a zero momentum 
particle.  As shown below  [Section 2.3(d)], we use this observation 
and minimize 
the total sum of energies of $N$ particles to prove that particles in 
the G-state of 
a SIB have equal energy as depicted by Fig.1(C)  rather than different 
energies 
as depicted by Fig.1(B) and concluded by SPB theories [1,2,6]  

\bigskip
\noindent
{\bf (d).}  Assuming that $i$-th atom exclusively occupies a cavity of 
volume 
$v_i$ with least possible energy, the ground state energy of the system 
can be written as 

$$E = \sum_i^N \frac{h^2}{8mv_i^{2/3}},  \eqno(4)$$ 

\noindent 
with the condition 
$$\quad \sum_i^N v_i = {\rm V}. \eqno(5)$$ 

\noindent 
A simple algebra using these two relations reveals that $E$ (in Eqn. 4) 
has its minimum value only if all $v_i$ are equal to $V/N$ and this 
renders 
$$E_o =  \frac{Nh^2}{8md^{2}} = N\varepsilon_o = N\frac{\hbar^2 q_o^2}{2m} 
 \eqno(6)$$ 

\noindent 
which implies $q_o = \pi/d$. Evidently, $E_o$(SPB) is also not ensured for 
the minimum 
value of the energy contributions from the relative motions of 
two particles.  It is evident that Eqns.(4) and (6) are consistent with an 
obvious condition known as excluded volume condition [13] which states 
that each particle such as $^4He$ atom occupies certain volume exclusively 
due to its short range HC interaction with other particles.   In formulating 
our theory [4,5] of a SIB like liquid $^4He$, we obtain Eqns.(4) and (6) 
by using our conclusion that two HC particles have to have $\lambda/2 \le d$
[14] ({\it i.e.}, $q \ge \pi/d$) which is consistent with excluded volume 
condition [13] as well as with wave packet manifestation of a particle 
(a consequence of wave particle duality). Not surprisingly, it appears 
that at low energy the size of wave packet $\lambda/2 = \pi/q = h/2\sqrt{2mE}$ 
serves as the effective size of a HC particle and, interestingly, 
$a \approx 100\AA$ of $^4He$ atoms at 1 mK energy and $a \approx 3\AA$ 
at about 1K seems to satisfy $E^{-1/2}$ dependence.

\bigskip
\centerline {\bf 3. Other important observations }

\bigskip 
As evident from [1, 10, 11], SPB theories replace two body 
{\it impenetrable} hard core (HC) potential defined by 
$V_{HC}(r_{ij} > a) = 0$ and $V_{HC}(r_{ij} \le a) = \infty$ by 
a {\it penetrable} $\delta-$potential when they apply perturbative 
method by using 
$$V_{HC}(r_{ij}) \equiv \frac{4\pi a\hbar^2}{m}\delta{(r_{ij})}. 
\eqno(7)$$ 

\noindent 
where right hand side potential term does not assume infinitely 
high value even for $r_{ij} = 0$ since the strength of Dirac delta 
({\it viz.}, ${4\pi a\hbar^2}/{m}$) [10, 11] is a small finite value. 
One may find that no simple logic justifies this equivalence of two 
physically different potentials, -one having infinitely high value 
for $r_{ij} = 0$, while the other having only a small finite value. 
Similarly, the variation approach based on Jastrow or 
Jastrow-Feenberg function [15, 16] ignores the implication of the 
shape independent approximation as stated in Section 2.3(a).        
 
\bigskip
\noindent
\centerline{\bf  4. Concluding remarks}  

\bigskip
\noindent 
(i) The ground state of a SIB [G-state(SPB)] as 
conculded by using single particle basis [1] is not a state of least 
possible energy. 

\bigskip
\noindent 
(ii) Subjecting every two particles in G-state(SPB) to the 
conditions for the minimization of their energy [ {\it viz.} with  
$|{\bf K}| = 0$ and Eqns.(4 and 5) for the energy of residual $q$ ], 
we obtain nothing but G-state(PPB).  This not only  underlines 
the accuracy of our theory [4,5] which renders G-state(PPB) but also 
concludes that one has to have {\it particle pair basis} as the 
starting point of the theory in question as we do in [4,5].

\bigskip
\noindent   
(iii) When the particles in a system interact through a two body 
interaction, a pair of particles forms its natural and logical basic unit.  
Naturally, theoretical models based on single particle basis can not 
be expected to reveal results of desired accuracy as shown here  
for G-state(SPB).  In view of the fact that the accuracy of our 
understanding of the G-state of a system plays an important role in 
revealing its physical behavior, the present analysis helps in 
establishing that $p=0$ condensate does not represent the true 
form of condensate responsible for the superfluidity of a SIB; 
as concluded in [4,5], true condensation occurs in a state of pair of 
particles each having $q=q_o$ and $K=0$ [4,5].  
 
\bigskip
\noindent
(iv)  In agreement with the results of this study, we find [4,5] that 
inter-particle 
HC repulsion decides crucially important G-state of a SIB like liquid 
$^4He$ because 
it serves as the origin of the excluded volume condition [13] which 
helps in 
determining the least possible energy of a particle, while 
inter-particle attraction 
is found to bind all particles into a single unit (a kind of macroscopic  
molecule) with a net binding energy which serves as the origin of 
several important 
aspects of superfluid SIB [4,5].  The G-state of a SIB has nothing 
but the potential energy $-V_o$ (representing the flat potential surface 
on which the particles are free to move in order of their locations [4.5]) 
added with the energy of zero-point motion which too 
serves as a potential energy responsible for the zero-point repulsion.  
It does not 
have different number of particles with different momenta 
(including $p=0$) 
as concluded by SPB theories and represented by Fig. 1(B); rather, as 
depicted in Fig.1(C), each particle has $p = \hbar q_o = h/2d$ 
momentum and cooresponding zero-point energy, $\varepsilon_o = h^2/8md^2$. 

\bigskip
\noindent
(v).  Even in case of a system of non-interacting bosons (SNIB) contained 
in a box of 
finite size, $L$, the G-state is represented exactly by the superposition 
of two plain 
waves of equal and opposite momenta (q, -q) [with $q = q_o = \pi/L$ which is 
not zero],  not by a plane 
wave which implies that each particle has non-zero momentum/energy 
(indeed of a infinitely small value for $L >> d$).  Evidently, there 
is no condensate of particles with $p=0$ and the G-state in its 
description is identical to that concluded in [4,5] for a SIB like 
liquid $^4He$.   Having a similar
analysis to the particles in a dilute bose gas trapped in a harmonic 
potential [17], one can easily 
find that $p=0$ condensate does not exist in the so called BEC state 
discovered in 1995 since here 
too the G-state has particles only with non-zero energy/ momentum.    

\bigskip
As such our present analysis unequivocally concludes that the popular 
belief {\it that there exists $p=0$ condensate in the superfluid phase 
of a SIB such as liquid $^4He$ and trapped dilute gases} is nothing but 
a myth and this underlines the potential of our theory [4,5] to explain 
the behavior of a SIB. 
.      

\bigskip
\noindent
{\it Acknowledgment :}  The author is thankful to Drs Bimalendu Deb (IACS Kolkata) and R. P. 
Bajpai (NEHU, Shilllong) for fruitful discussions.

\newpage

 \bigskip
\begin{figure}
\begin{center}
\includegraphics[angle = 0, width =.6\textwidth]{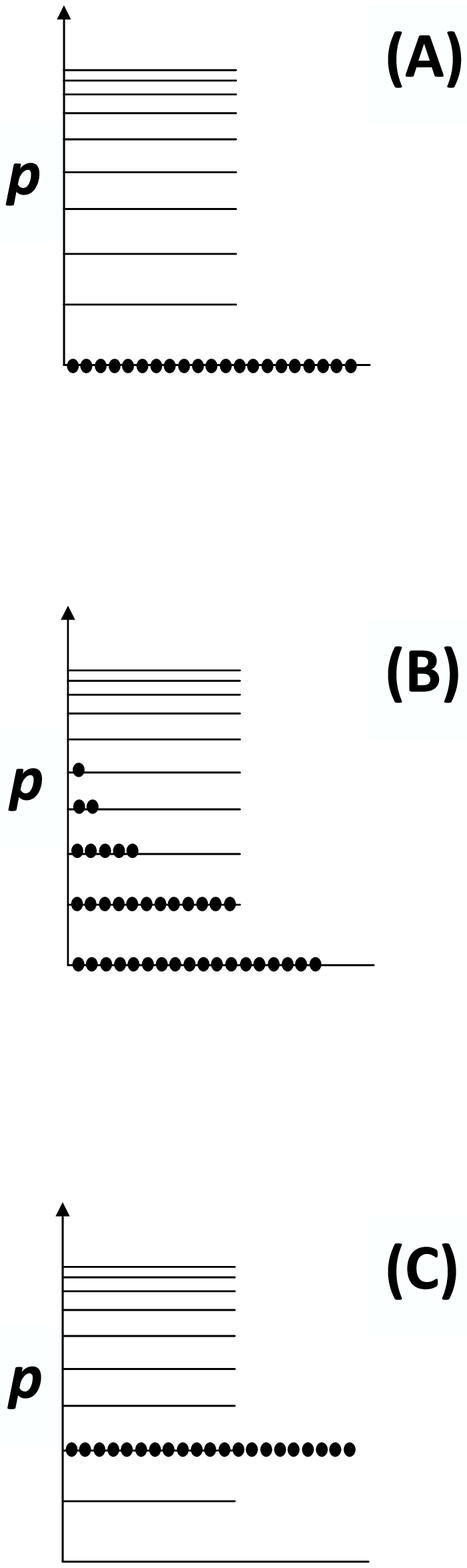}

\end{center}
\bigskip
\noindent
Fig.1 : Schematic of distribution of $N$ bosons in their 
ground state. (A) All the $N$ particles occupy $p=0$ state 
in a system of non-interacting bosons, (B) depletion of 
$p=0$ condensate ({\it i.e.} only a fraction of $N$ occupy 
$p=0$ state) in weakly interacting boson system as predicted 
by Bogoliubov model [6], and (C) all the $N$ particles occupy 
a state of $q = \pi/d$ and $K=0$ as concluded from our 
recent theory [4,5].

\end{figure}

\newpage
\begin{thebibliography}{??}  
\bibitem[1]{Kn:gnus}
Anderson, J.O.: Rev. Mod. Phys. 76, 599-639 (2004).
\bibitem[2]{Kn:gnus}
Yukalov, V.I., Kleinert, H.: Phys. Rev. A73, 063612(2006).
\bibitem[3]{Kn:gnus}
Ettouhami, A.M.: Reexamining Bogoliubov's theory of an 
interacting Bose gas,  arXiv:cond-mat/ 0703498v1 (2007).
\bibitem[4]{Kn:gnus}
Jain, Y.S.: Macro-orbitals and microscopic theory of a system 
of interacting bosons, arXiv:cond-mat/0606571.  
\bibitem[5]{Kn:gnus}
Jain, Y.S.: New approach to the microscopic theory of a system of
interacting bosons -I. Basic foundations and Superfluidity (sent 
for publication) is a revised version of [4] with additional 
sections); its pdf copy can be requested from the author.   
\bibitem[6]{Kn:gnus}
Bogoliubov, N. N.: J. Phys, USSR, 11, 23-32 (1947).  
\bibitem[7]{Kn:gnus}
Pashitskii, E.A., Mashkevich, S.V., Vil'chinskii, S.I. Phys. Rev. Lett. : 89, 075301 (2002); Tomchenko, 
M.: Low Temp. Phys, 32, 38 (2006),
\bibitem[8]{Kn:gnus}
Chutia, S.: A study of certain properties of superfluid helium-4 based 
on macro-orbital theory, Ph.D. Thesis (2007), Department of Physics, 
North-Eastern Hill University, Shillong - 793 022, India.   
\bibitem[9]{Kn:gnus}
Fixing the positions of $N$ particles demands fixation of only $N$ 
points, while fixing of $N(N-1)/2$ CM points fixes more than desired 
number of points related to the positions  
of $N$ particles.  Since $N(N-1)/2 \ge N$ is true for all $N \ge 3$, 
fixing of $N(N-1)/2$ CM points naturally ensures fixation of the 
positions of $N$ particles except for a system of $N = 2$ which has 
only 1 CM point;  this rightly agrees with the fact 
that one can not fix the positions of two particles or their 6 degrees 
of freedom by fixing the position of only 1 CM point, -equivalent to 
fixing only 3 degrees of freedom leaving 3 degrees of freedom to remain 
unfixed and these are two rotations and one vibration.  We also note 
that the coordinates of each CM point are related 
to the position coordinates of two chosen particles, positions of all 
the $N(N-1)/2$ CM points get automatically fixed if we fix the positions 
of $N$ particles.     
\bibitem[10]{Kn:gnus}
Pathria, R.K.: Statistical Mechanics,
Pergamon Press Oxford (1976). 
\bibitem[11]{Kn:gnus}
Huang, K.: Statistical Mechanics, Wiley Eastern Limited,
New Delhi (1991).
\bibitem[12]{Kn:gnus}
Grisenti et. al.: Phys. Rev. Lett.  85, 2284-2287 (2000) .
\bibitem[13]{Kn:gnus}
Kleban, P.: Phys. Letters A 49, 19-20(1273).
\bibitem[14]{Kn:gnus}
Jain, Y.S. : Cent. Euro. J. Phys. 2, 709-719(2004).
\bibitem[15]{Kn:gnus}
C. -W. Woo in The Physics of liquid and solid helium, K.H. Benneman and J.B. Ketterson (Editors), 
Part-I, Wiley, New York (1976).  
\bibitem[16]{Kn:gnus}
Feenberg, F.: The theory of Quantum Fluids, Academic, New York, 1969).
\bibitem[17]{Kn:gnus}
Dalfovo, F., Giorgini, S. Pitaevskii, L.P. and Stringari, S.: Rev. Mod. Phys. 71 463-512 (1999); 
arXiv/Cond-mat/9806038.
 
     

\end{thebibliography}
\end{document}